\documentclass{aa}
\usepackage{txfonts}
\usepackage{graphicx}

\begin{document}

  \title{
          XMM-Newton EPIC Observation of the Galaxy Cluster A3667
  \thanks{
          Based on observations with XMM-Newton, an ESA Science Mission
          with instruments and contributions directly funded by ESA Member
          States and the USA (NASA)}
   }

\author{Ulrich G. Briel,\inst{1} 
        Alexis Finoguenov,\inst{1} and 
        J. Patrick Henry\inst{1,2} 
        }

  \offprints{U. G. Briel, \email{ugb@mpe.mpg.de}}

 \institute{Max-Planck-Institut f\"ur extraterrestrische Physik,
            85740 Garching, Germany
\and
            Institute for Astronomy,
            2680 Woodlawn Drive, Honolulu, Hawaii 96822,  USA}
   \date{Received 5 December 2003 / Accepted 4 June 2004}

\authorrunning{Briel et al.}

\titlerunning{XMM-Newton EPIC Observation of the Galaxy Cluster A3667 }

%%%%%%%%%%%%%%%%% BEGINN ABSTRACT %%%%%%%%%%%%%%%%%%%%%%%%%%%%%%%%

\abstract{ The Abell cluster of galaxies A3667 was observed with
XMM-Newton in 6 partially overlapping pointings. We present here the
resulting X-ray surface brightness, temperature, entropy and pressure
maps and discuss the structure of this cluster on scales from 0.1 to
30 arcminutes. Based on these observations we refine the origin of the
A3667 cold front to the displacement of the low entropy, high metal
abundance gas from the current pressure peak of the cluster. We argue
that the mushroom shape of the cold front observed here is similar
to what is seen in some numerical simulations. We also present the first
evidence for a dynamically significant angular momentum in the cold
front.
}

%%%%%%%%%%%%%%%%% END ABSTRACT %%%%%%%%%%%%%%%%%%%%%%%%%%%%%%%%

   \maketitle

%%%%%%%%%%%%% BEGINN INTRODUCTION %%%%%%%%%%%%%%%%%

\section{Introduction}

\begin{table*}[t]
\caption[ ]{Journal of Observations}
%\begin{flushleft}
\begin{tabular}{ccccccccc}
\hline
\rule{0mm}{4mm}
%\\[0.5ex]
Date & Name of     & RA(2000) & DEC(2000) &  \multicolumn{2}{c}{performed/effective exposures} & XSA ID & Orbit\\
     & Observation &          &           & pn (ksec) & 
\multicolumn{1}{c}{MOS1--2 (ksec)} &&\\
\hline
\rule{0mm}{4mm}
2000 Sep 9   & A3667 f1 & 20 13 03.5 & -56 53 00 & 14.8/2.4  & 16.5/5.7 - 16.3/5.1  &0105260101&144\\
2000 Oct 13  & A3667 f2 & 20 11 08.8 & -56 38 29 & 15.0/7.6  & 18.9/0.4 - 18.9/0.8  &0105260201&155\\ 
2000 Oct 3   & A3667 f3 & 20 13 07.0 & -56 43 08 & 13.1/13.1 & 17.0/17.0 - 17.0/17.0&0105260301&150\\
2000 Oct 3   & A3667 f4 & 20 12 12.5 & -56 37 15 & 12.7/12.7 & 16.6/16.6 - 16.6/16.5&0105260401&150\\
2000 Oct 4   & A3667 f5 & 20 10 55.7 & -56 48 06 & 13.1/13.1 & 13.1/11.6 - 13.1/11.8&0105260501&150\\
2000 Oct 2   & A3667 f6 & 20 11 53.2 & -56 54 45 & 21.6/18.8 & 25.5/23.8 - 25.5/23.7&0105260601&149\\ 
\hline
\end{tabular}
%\end{flushleft}
\end{table*}

When the Abell Cluster A3667 was observed with ROSAT (Knopp et
al. \cite{gk96}), the X-ray image revealed a surface brightness of varying
ellipticity and a surface brightness discontinuity SE from the core. The
counting statistics of the ROSAT observation was too low to obtain a
temperature map, but it was concluded that the cluster is not in a relaxed
configuration but rather in the process of merging since the fit to a
constant temperature was rejected at 95\% confidence. Evidence was found for
excess emission towards the NW, apparently coincident with a group of
galaxies.

Analysis of Chandra observations by Vikhlinin et al. (\cite{av01-1}
and \cite{av01-2}) made A3667 a definition of a new phenomenon, a
contact discontinuity or a {\it cold front}, associated with the
transonic motion of the cluster core.  In the follow-up paper, a
hydrodynamic instability was identified perpendicular to the cold
front (Mazzotta et al. \cite{pm02}) pointing to the transient nature
of the cold front, as it is gradually disrupted via Kelvin-Helmholtz
instability.

Since the discovery of the cold front, a number of numerical
experiments have identified similar features observed in the process
of cluster merging.  Further, Markevitch et al. (\cite{mm03}) have shown 
that 70\% of Chandra images show sharp edges, indicating a wide-spread 
occurrence of this phenomenon.

Important insight on the possible formation of cold fronts has been
given by the numerical simulations of Heinz et al. (\cite{sh03}),
where it has been suggested that the cold front is produced by the
upstream motion of low entropy gas from the minimum of the
gravitational potential of the merging subcluster.  The low
temperature of the front was produced even when the initial state of
the gas was isothermal, with cooling due to adiabatic expansion.
Other simulations (e.g. Ricker \& Sarazin \cite{rs01}), suggest an
important role of the angular momentum initially possessed by the gas
associated with the merger of two cluster cores.

In this {\it Paper} we present the result of 6 partially overlapping
observations of the Abell cluster A3667 with the EPIC detectors on
board XMM-Newton. In Section 2 we describe the observations and
present the data reduction.  In Section 3 we discuss the structure of
the X-ray surface brightness distribution on large scales, the
hardness ratio and temperature map of the cluster as well as deduced
pressure and entropy maps. We confirm some of the features of the cold
front suggested by Heinz et al. (\cite{sh03}), and also identify the
features indicating the presence of significant angular momentum in
the cold front.

A3667 is at a redshift of 0.053. With a Hubble constant of 70 km sec$^{-1}$
Mpc$^{-1}$, and $\Omega_M=1-\Omega_\Lambda=0.3$, the luminosity distance is
237 Mpc and the angular scale is 62 kpc per arcminute.

%%%%%%%%%%%%% END INTRODUCTION %%%%%%%%%%%%%%%%%

%%%%%%%%%%%%% BEGIN OBSERVATIONS AND DATA REDUCTION %%%%%%%%%%%%%%%%%

\section{Observations and Data Reduction}

The region of the Abell cluster of galaxies A3667 was observed by the
XMM-Newton Satellite (Jansen et al. \cite{fj01}) as part of the GTO program 
of the XMM telescope scientist. Table 1 shows the journal of the
observations with information on the pointing directions and observing
times. 

In the analysis that follows we will present the data obtained with the two
MOS-CCD cameras (Turner et al. \cite{mt01}), operated in their full-frame
mode, and with the pn-CCD camera (Str\"uder et al. \cite{ls01}), operated 
in the extended full-frame mode. See the above papers for a description 
of the different modes. All cameras used their medium filters to block
the visible light. 

\begin{figure*}
\includegraphics[width=17cm]{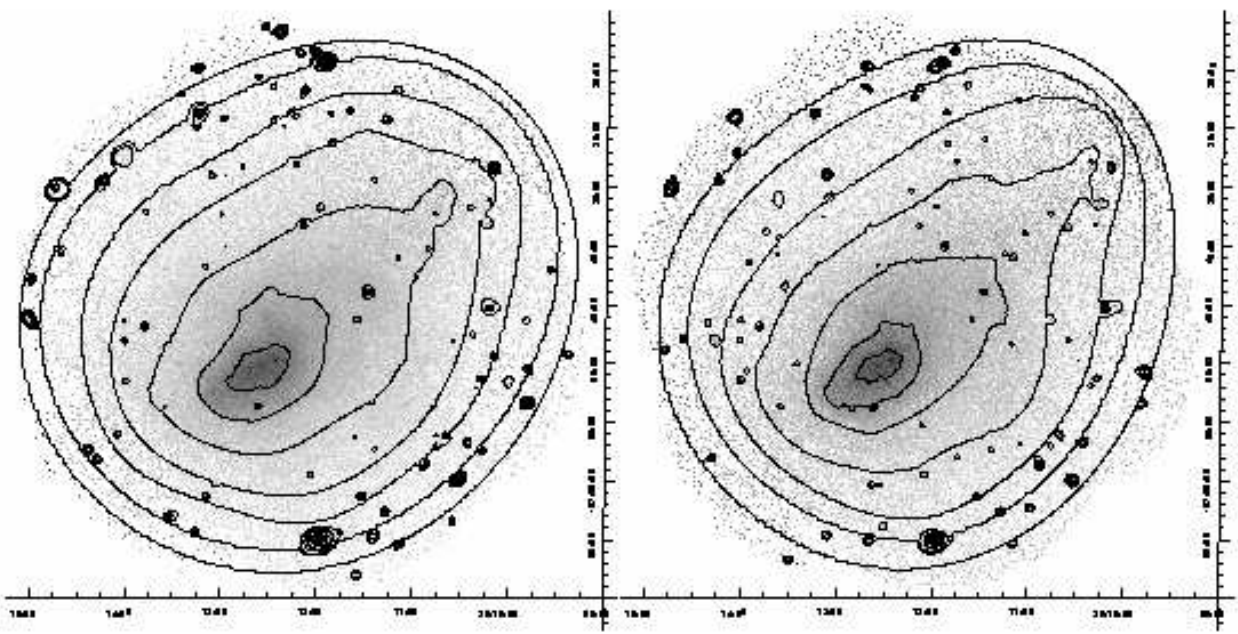} 

\caption{The merged EPIC image of the Abell cluster of galaxies A3667 in
the 0.8 to 2.0 keV (left panel) and in the 2.0 to 7.5 keV (right panel)
energy bands. Contours indicate the levels of equal intensity in the
wavelet-decomposed image, spaced by a factor of 3 starting at
$6.8\times10^{-5}$ MOS1 cnts s$^{-1}$ arcminute$^{-2}$ level.  Coordinate
grids are shown for the J2000.
\label{f:raw}}

\end{figure*}

\begin{figure*}
\includegraphics[width=17cm]{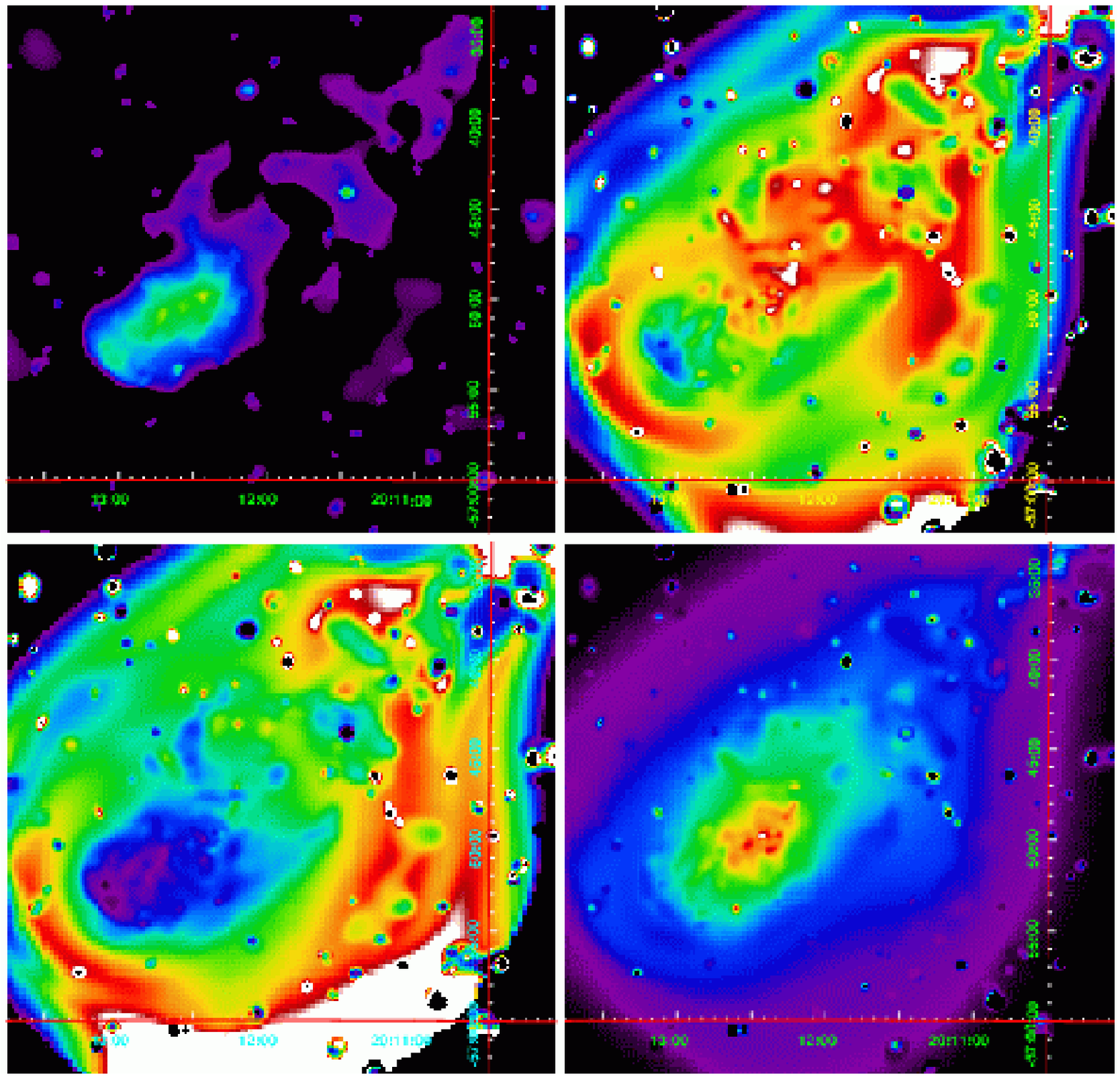} 

\includegraphics[width=17cm]{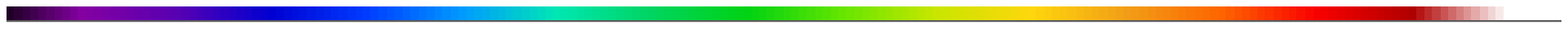}

\caption{Upper left: wavelet-reconstructed image of A3667 in
the 0.8--2 keV band ($I$). To highlight the image structure, only the
smallest wavelet scales are shown. Upper right: temperature map ($T$),
deduced from the wavelet-filtered ratios between 0.8--2 and 2--7.5 keV
images. Lower row: pseudo entropy (left) and pressure (right) maps, derived
as $T / \sqrt[3]{I}$ and $T\times \sqrt{I}$. The color legend for the
four images is as follows: violet, blue, green, orange, red, white
corresponds to: surface brightness 0.001, 0.01, 0.04, 0.07, 0.16, 0.2 in MOS1
cnts s$^{-1}$ arcminute$^{-2}$; temperature 3 keV, 4 keV, 5 keV, 6 keV, 7
keV, $\ge$ 8 keV; pseudo entropy 50, 100, 130, 170, 200, $>250$ in arbitrary
units; pseudo pressure 0.01, 0.05, 0.09, 0.12, 0.15, 0.20 in arbitrary
units.
\label{f:hr}}
\end{figure*}

\begin{figure*}
\includegraphics[width=17cm]{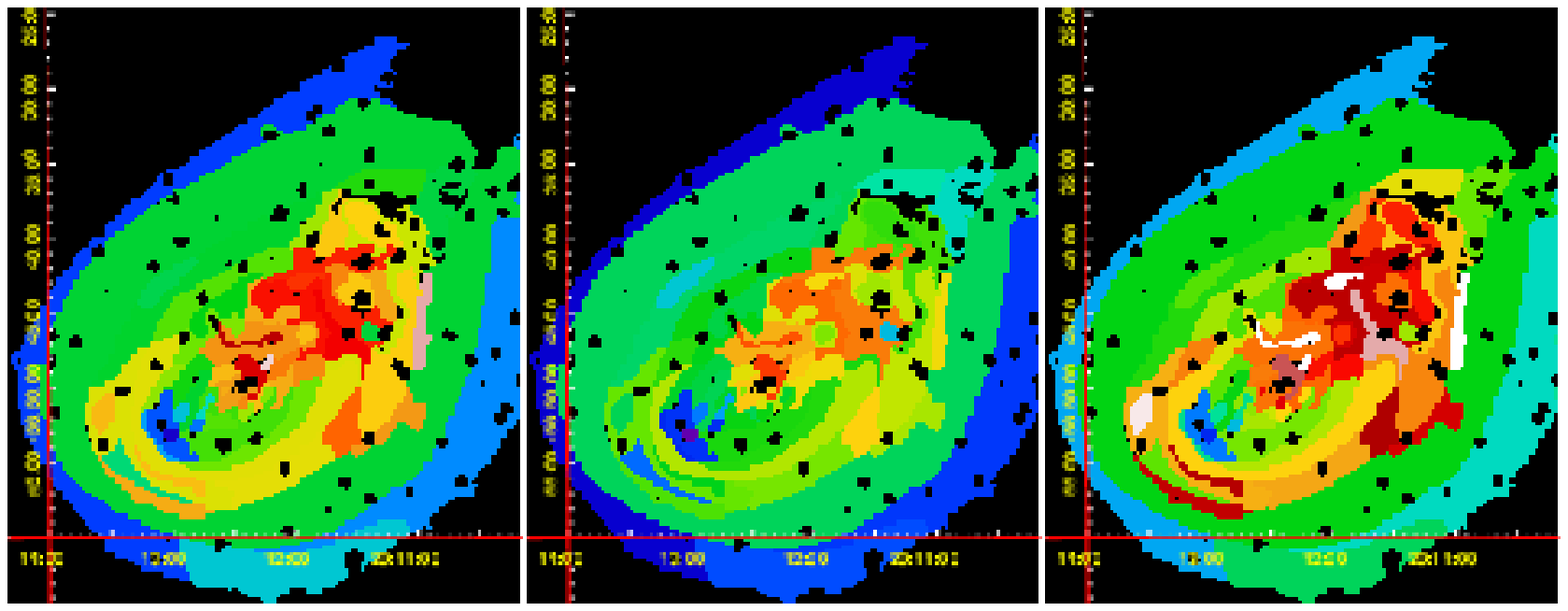}

\includegraphics[width=17cm]{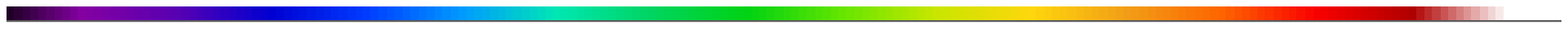}
\caption{Temperature map (left), derived from the spectral analysis
with 1$\sigma$ lower (middle) and upper (right) limits on the
temperature, displaced in the same color scale. Blue is 4 keV, green is
5 keV, orange is 6 keV and red is 7 keV, white is 8 keV.
\label{f:te}}
\end{figure*}

\begin{figure*}
\includegraphics[width=17cm]{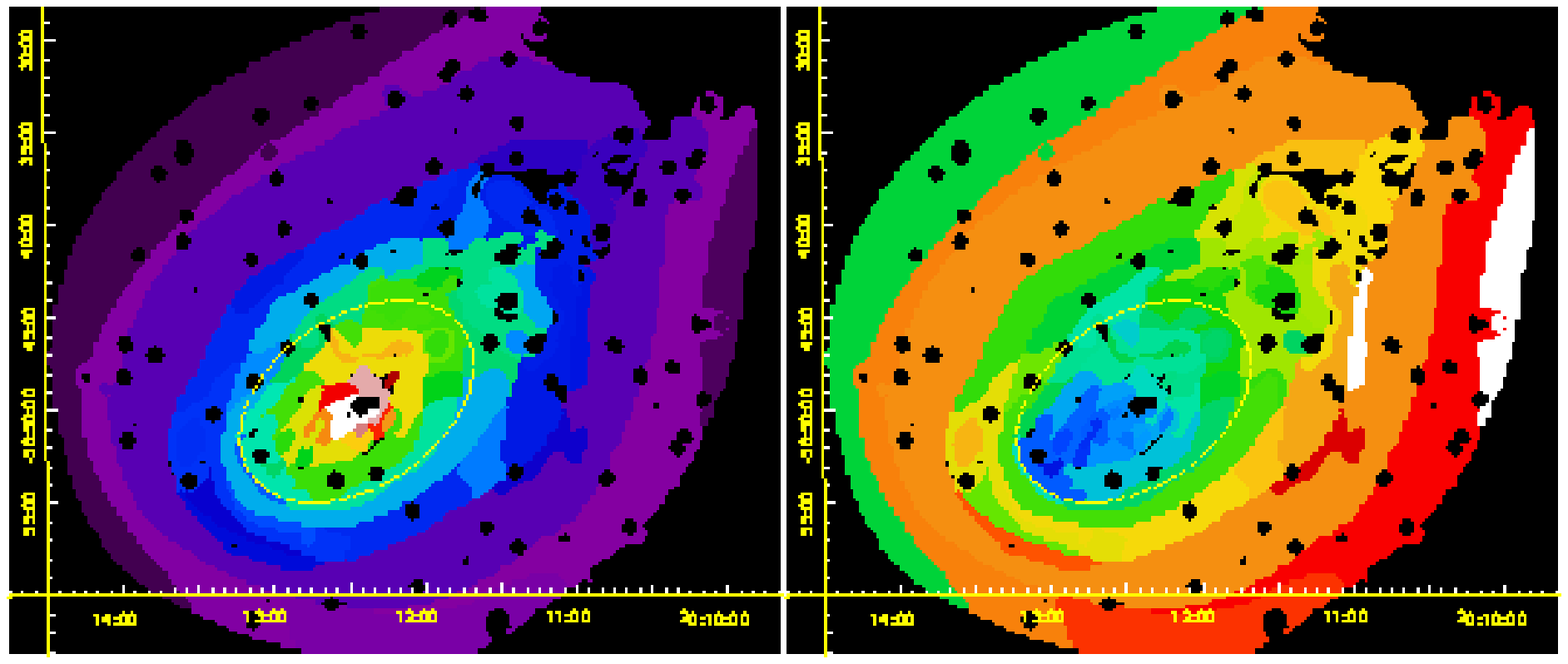}

\includegraphics[width=17cm]{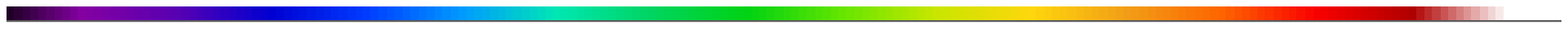} 
\caption{Projected pressure and entropy maps of A3667, derived from the
  spectral analysis. A yellow ellipse has a center at RA = 20:12:27.4 and
  Dec = -56:49:34.5, a major axis of 417.03$^{\prime\prime}$ and a minor
  axis of 275.92$^{\prime\prime}$ and with a positional angle of $34^\circ$,
  denoting the pressure core of the A3667. The color legend for the two
images is as follows: violet, blue, green, orange, red, white corresponds to:
entropy 100, 150, 350, 500, 600, 800 in arbitrary units; pressure 0.01,
0.05, 0.10, 0.15, 0.20, 0.3 in arbitrary units.
\label{f:sp}}
\end{figure*}

\subsection{Merged image production}

Our goal here is to produce soft and hard images merging all EPIC
detectors and pointings. Initial steps of data reduction were
performed using XMMSAS 5.3.  During the observations the proton
induced background often varied, and in flares reached count rates up
to several hundred counts/sec in the energy band from 0.15 to 10
keV. To increase the signal-to-noise ratio and also to minimize the
problems of de-vignetting of residual proton induced events, we
selected the good time intervals where the detector countrate in the
10--15 keV band did not exceed a given threshold. While in our
previous investigations we have used a fixed threshold (e.g. Briel et
al. \cite{ub01}), in this work we have adopted the approach based on
the analysis of the countrate histogram, as described in Zhang et
al. (\cite{zh03}).  This method is more sensitive to the background
conditions during the observations and has now been adopted in the
analysis of other groups (e.g. De Luca \& Molendi \cite{lm04})
The resulting net-observing times are given in Table 1 for
the different instruments. As can be seen in that table, the fields f1
and f2 are heavily contaminated by proton flares.

With these screened photon event files we produced MOS and pn images of the
individual observations in the energy bands from 0.8 to 2.0 and 2.0 to 7.5
keV. To reduce the widths of the gaps in the pn-images, we included photons
near the pn-CCD borders, near bad pixels, and near offset columns. This can
be done if one is only interested in producing qualitative images in wide
energy bands. For the spectral fitting, described  in \S\ref{s:spe}, we 
exclude all those border photons, since a small fraction of those photons 
have the wrong energy, typically due to the registration of a double event 
pattern as a single event. 

The XMM-Newton mosaic observation of A3667 covers a circle of about
$24^\prime$ radius centered on RA=303.02862, Decl=-56.75183 (J2000).
Merging the individual observations together have shown that the
signal was above the background only in the central $11^\prime$ of the
cluster due to high level of the non X-ray background. As this leaves
3/4 of the mosaic out, we have used the outermost parts of the mosaic
for the in-field background estimates (see e.g. Zhang et al.
\cite{zh03} and Finoguenov et al. \cite{af03-1} for the details) to
finally extend the imaging to $21^\prime$.

We assumed that the total background for each observation suffered no
vignetting. This assumption is approximately correct for the non X-ray
instrumental background as well as the soft proton flares (Lumb et
al. \cite{dl02}). It is not true of the unresolved X-ray
background. However, background subtraction is most important for the hard
(2--7.5 keV) band where the unresolved X-ray background is a few percent of
the total. Thus the error coming from assuming the X-ray background is
unvignetted is comparable to the 4\% error introduced by the assumption of
no vignetting in the soft proton component (Lumb et al. \cite{dl02}).

We generated the flat background maps using exposure maps with no vignetting
information (also known as chip masks), normalized them separately for each
observation and instrument to the observed background level using data
outside the central $21^\prime$. We avoided point sources in selecting the
regions for the background normalization. After subtracting these background
maps from each image, we corrected the residual for vignetting and exposure
using the latest calibrations (Lumb et al. \cite{dl03}), which have
been incorporated within the XMMSAS 6.0 release. Finally we combined the MOS
and pn data by normalizing the exposure maps by a factor of 2.4 to account
for the overall difference in the sensitivity between the MOS and pn. We
also correct the MOS hard band images by an additional factor of 1.05. These
corrections yield the same MOS and pn hardness ratios for the same
temperature in the 2--12 keV range to better than 0.5 keV.

The resulting merged pn plus MOS background subtracted and exposure
corrected images in the 0.8--2 and 2--7.5 keV bands are shown in
Fig.\ref{f:raw}. The images are devoid of any detector structure because
we used all available photons near CCD borders, bad pixels and columns
as well merging different pointings and instruments. The overlaid
contours in units of MOS1 counts s$^{-1}$ arcminute$^{-2}$ are from
the wavelet decomposition of the same image, as described below.

\subsection{Wavelets analysis}

The broad-band images can be used for making visible intensity
structures, and also variations of the temperature of the X-ray
emitting plasma by producing hardness-ratio maps. In addition, one can
produce pressure and entropy maps of the plasma by combining the
surface brightness map and the hardness ratio map. Useful hardness
ratio maps can only be produced from smoothed surface brightness
maps. There exists a variety of different smoothing procedures like
top-hat smoothing, Gaussian smoothing or adaptive smoothing (e.g.
Churazov et al. \cite{ec99}). We applied the wavelet decomposition
method, which is described in detail by Vikhlinin et
al. (\cite{av98}).  The advantage of using wavelets consists in
background removal by spatial filtering and a control over the
statistical significance of the detected structure. Complications
arise due to splitting the image into discrete scales, which we
overcome by additional smoothing applied before producing the hardness
ratio map. Use of wavelets provide us with a quick and dirty method to
identify the regions susceptible to temperature variations.  Another
important feature is the high spatial resolution achieved in detecting
the small-scale structure, as wavelets do not smear the data.

We show the results of the broad-band image investigation in
Fig.\ref{f:hr}. In the top left panel we show the small-scale structure
detected in the 0.8--2 keV image, using the wavelet decomposition method.
The top right is the hardness of the emission deduced from the ratio of the
wavelet-reconstructed images in the 0.8--2 and 2--7.5 keV bands.  Using
a calibrated template, the hardness can be interpreted as temperature, as
given in the figure caption. The lower left and right images are the
projected pressure and entropy maps respectively.  They are constructed
using the wavelet-smoothed surface-brightness map in the 0.8--2 keV energy
range as an indicator of the electron density squared and the hardness ratio
map as temperature distribution, and using the definitions of the pressure
$P\sim T\times \sqrt{I}$ and of the entropy $S\sim T / \sqrt[3]{I}$. 
There is an important difference between bolometric X-ray luminosity and a
luminosity of the nearby clusters in the soft band (typically ranging from
0.5 or so to 2 keV), that is used in deriving the formulae for pseudo
pressure and entropy. The difference stems from specifics of the cooling
function for clusters hotter than 2 keV.  While bolometric luminosity
depends on $n_e^2 \sqrt{T}$, our calculations for both pn and MOS show that
luminosity in the soft band is to a 10\% precision constant with changing
temperature in the 2--15 keV range and is actually decreasing with
increasing temperature. B\"ohringer (1996) first pointed out this
temperature insensitivity of the conversion between soft band counts to flux
for nearby clusters in application to the ROSAT PSPC data. He also mentioned
that absence of the temperature dependence is valid for a wide range of
galactic $nH$ values.

\subsection{Spectral analysis}\label{s:spe}

The spectral analysis was carried out using only the pn
data. Background subtraction here is more demanding than for the
imaging analysis since the spectrum of the background must be
estimated, not just its overall level. Therefore we have investigated
several different background accumulations, one by Read \& Ponman 
(\cite{ar03}) and others performed later in the XMM mission (e.g. 
APM08279+5255; Hubble Deep Field South). We find that for the A3667 
observation the detector background in the 10--15 keV range corresponds 
closest to the accumulations of Read \& Ponman (\cite{ar03}), which 
consists of several observations performed around the observing dates 
of the A3667 mosaic. Hence for the following analysis we used the Read 
background.

A first check was to reproduce an integral temperature and compare it
with values measured by other X-ray instruments. With ROSAT, Knopp et al.
(\cite{gk96}) found a temperature of 6.3 +0.5/-0.6 keV
in a circle with a radius of $20^\prime$ centered on the peak
emission. Using ASCA, Markevitch et al. (\cite{mm98}) found a
temperature of 7.0 $\pm$0.6 keV in the same region. As the cluster
exhibits temperature structure, we find the XMM derived integral
temperature inside the $20^\prime$ radius to depend on the energy
range selected for the analysis. From our data, we find a temperature
of $6.1\pm0.1$ keV for the 0.4--10 keV range and $6.4\pm0.1$ keV if we
set the low energy threshold in the 0.5 to 1 keV range, but both fits
are at an unacceptable $\chi^2$.  Acceptable values of $\chi^2$ are
obtained with low energy cuts of 1.5 keV or above, with the resulting
temperature of $7.1\pm0.1$ being no longer a function of the selected
energy range. This resulting temperature is in very good agreement
with the ROSAT and ASCA measurement. We also find element abundances
of Si of $0.60\pm0.09$ solar, S of $0.20\pm0.11$ solar and Fe of
$0.23\pm0.01$ solar (using the photospheric values from Anders \&
Grevesse \cite{ea89} as a reference). These values agree with the
results on the element abundance in clusters obtained by ASCA
(e.g. Finoguenov et al. \cite{af00} and \cite{af03-2}; Baumgartner et
al.  \cite{wb03}) and confirm both the prevalence of SN II in the
enrichment and a reduced S yield in SN II, as found in the above ASCA
studies.

To determine a temperature map from the XMM-Newton mosaic, we used the
surface brightness and hardness ratio map, described in the previous
section, to define independent regions for spectral fitting.  A mask file
was built by using the changes in the hardness ratio that correspond to
temperature in the ranges
3.0-3.5-3.8-4.0-4.2-4.4-4.8-5.2-5.6-6.0-6.5-7.0-7.5-8-9-10 keV and which
have equal intensity within a factor of two. Taking the isolated regions of
equal temperature and intensity separately and imposing the criteria that
the regions should be larger than the PSF width ($15^{\prime\prime}$) and
contain more than 1000 counts in the raw pn image, we obtain the final mask
used to extract the photons for the spectral fitting. That procedure
led to 70 regions for which we performed the spectral analysis.

For the spectral fits, we have chosen a single-temperature APEC
model. Spectral analysis of the regions of low surface brightness
reveals the presence of an excess soft X-ray emission at energies
below 0.7 keV. This component is not required in the regions of high
X-ray intensity, such as the cold front, but causes systematic
differences in the derived temperature in the single-temperature fits
as a function of selected energy band otherwise. A proper analysis of
this component requires a detailed understanding of the background
accumulation, such as that performed in Finoguenov et al.
(\cite{af03-1}), and is beyond the scope of the present paper. We
have excluded the effect of the soft component by selecting the energy
band starting at 0.75 keV. The upper energy boundary is determined by
strong instrumental background above the 7.9 keV. For selected
regions, which do not suffer from background problems mentioned above,
we have also performed a spectral analysis in the 0.4-10 keV band and
have confirmed our results obtained in the 0.75 to 7.9 keV range.

The results of the fitting procedure are shown in Fig. \ref{f:te}. Here we
show in the left panel the temperature map and in the middle and right panel
the lower and upper 1$\sigma$ limits of the temperatures respectively.  We
used the fitted temperature map in combination with the XSPEC APEC
normalization factor K, which is a measure of the surface-brightness, to
produce the projected pressure and entropy maps.  A more detailed
description of this procedure is given in Henry et al.  (\cite{ph04}).
Both pressure and entropy maps are shown in Fig. \ref{f:sp}.

\subsection{Hardness ratio vs temperature map}

In the selection of zones for the spectral analysis we have aimed to 
confirm the main features seen in the hardness ratio map. To these 
belong the cold front, the region in front of the cold front to the 
south-east, the material in which the cold front is embedded, and also
the region of the tail behind the cold front to the north-west. 
With a small number of zones on a half an
arcminute scale, most of the information is obtained on a few arcminute
scale inside the central $11^\prime$. At outer radii the structure is
analyzed on the $4-10^\prime$ scale.

Although the methods involved in producing the hardness ratio and
temperature maps are very different, the results are very similar.
Features seen in both with nearly the same morphology are the cold
front and its immediate environment as well as most of the temperature
variations in the tail.  Although some of the temperature in the
outskirts are confirmed, the scatter between the fitted temperatures
and the temperatures deduced from hardness ratio estimate
increases. The reason for that is probably the extrapolation of the
cluster shape on the largest scale by wavelets into the zones of low
signal-to-noise ratio. To illustrate that we point out that part of the
outermost wavelet contour in Fig.\ref{f:raw} passes though an
unobserved region.

\begin{figure*}
\includegraphics[width=17cm]{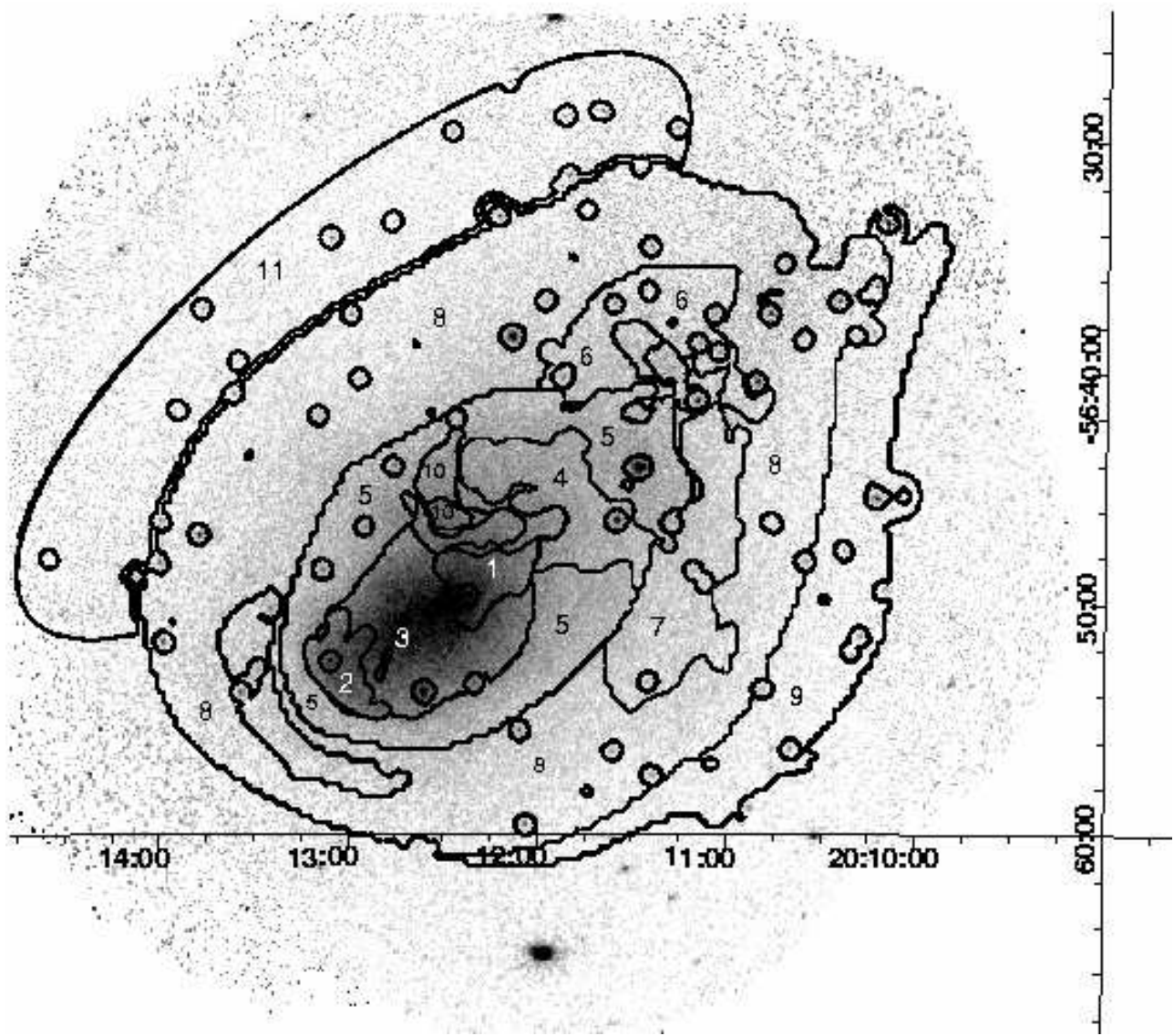}

\caption{Location of the main regions of A3667 on the image. The numbers
  correspond to the entries in Tab.\ref{t:spe}. Coordinate
grids are shown for the J2000.
\label{f:lab}}
\end{figure*}

\begin{table*}[ht]
{
\begin{center}
\footnotesize
{\renewcommand{\arraystretch}{0.9}\renewcommand{\tabcolsep}{0.12cm}
\caption{\footnotesize
Properties of main regions of A3667. 
\label{t:spe}}

\begin{tabular}{cccccccccccccc}
 \hline
 \hline
N & Name &$kT$ & $Fe/Fe_\odot$ & $norm$ & $\chi^2$ & N & $n_e$ & S
 & P, $10^{-12}$ & $M_{\rm gas}$ & $r_{min}$ & $r_{max}$\\
  &  & keV     &              & $10^{-3}$  & & d.o.f. & $10^{-4}$ cm$^{-3}$& keV cm
$^2$ & erg cm$^{-3}$ & $10^{12} M_\odot$ & kpc & kpc  \\
 \hline
1&nkT peak  &$7.1\pm0.2$&$0.27\pm0.03$&$ 5.38\pm0.05$&1.16&400&$24.9\pm0.1$&$ 387\pm 9$&$28.4\pm0.7$&$1.32\pm0.01$& 26& 281\\
2&cold front&$4.2\pm0.1$&$0.55\pm0.05$&$ 2.82\pm0.07$&1.31&152&$20.6\pm0.3$&$ 259\pm 6$&$13.9\pm0.3$&$0.83\pm0.01$&236& 439\\
3&core      &$5.8\pm0.1$&$0.33\pm0.02$&$14.85\pm0.07$&1.40&599&$23.6\pm0.1$&$ 325\pm 4$&$21.8\pm0.3$&$3.85\pm0.01$& 27& 360\\
4&hot tail  &$7.1\pm0.2$&$0.20\pm0.03$&$ 3.97\pm0.05$&1.25&337&$11.8\pm0.1$&$ 635\pm19$&$13.4\pm0.4$&$2.06\pm0.01$&224& 563\\
5&main-1    &$6.0\pm0.1$&$0.21\pm0.02$&$13.85\pm0.07$&1.41&623&$ 8.5\pm0.0$&$ 664\pm10$&$ 8.1\pm0.1$&$9.93\pm0.03$&208& 894\\
6&tail-N    &$6.2\pm0.3$&$0.19\pm0.07$&$ 1.84\pm0.06$&1.51&118&$ 5.4\pm0.1$&$ 931\pm50$&$ 5.3\pm0.3$&$2.08\pm0.03$&603&1167\\
7&tail-S    &$6.6\pm0.2$&$0.15\pm0.05$&$ 2.42\pm0.05$&1.24&177&$ 6.1\pm0.1$&$ 909\pm35$&$ 6.4\pm0.3$&$2.41\pm0.03$&424& 905\\
8&main-2    &$5.6\pm0.1$&$0.20\pm0.03$&$15.60\pm0.11$&1.46&656&$ 3.8\pm0.0$&$1066\pm23$&$3.4\pm0.1$&$25.29\pm0.09$&349&1492\\
9&main-3 S  &$4.5\pm0.3$&$0.10\pm0.09$&$ 1.89\pm0.10$&1.68&128&$ 2.5\pm0.1$&$1120\pm84$&$ 1.8\pm0.1$&$4.60\pm0.12$&680&1652\\
10&shear bar&$6.0\pm0.2$&$0.33\pm0.06$&$ 1.44\pm0.04$&0.85&114&$15.4\pm0.2$&$ 453\pm18$&$14.8\pm0.6$&$0.57\pm0.01$&166& 456\\
11&main-3 N&$4.0\pm0.3$&$0.36\pm0.13$&$ 2.69\pm0.21$&1.07&114&$ 2.4\pm0.1$&$1053\pm76$&$ 1.5\pm0.1$&$6.96\pm0.26$&823&1549\\
\hline
\end{tabular}
}
\end{center}
}
\end{table*}

%******************************************************************************

\section{Discussion}

%******************************************************************************

\subsection{Basic properties of A3667}

In order to tabulate the basic properties of A3667 we combined together
adjacent spectral regions having the same temperature to within 1 keV. We
list the properties obtained this way in Tab.\ref{t:spe} with their
$\pm1\sigma$ errors for one parameter.  Col. (1) labels the region according
to Fig.\ref{f:lab}, with the name of the region reported in (2), (3) lists
temperature in keV, (4) iron abundance as a fraction of solar photospheric
value of Anders \& Grevesse (1989), (5) XSPEC normalization, (6--7)
statistical quality of the fit. Derived quantities, that use an estimation
of the projected length, as described below are reported in
cols.(8--11). These are electron density, entropy, pressure and the (local)
gas mass. No account for the gas mass not associated with the directly
observed component was attempted.

\begin{figure*}
\includegraphics[width=8.0cm]{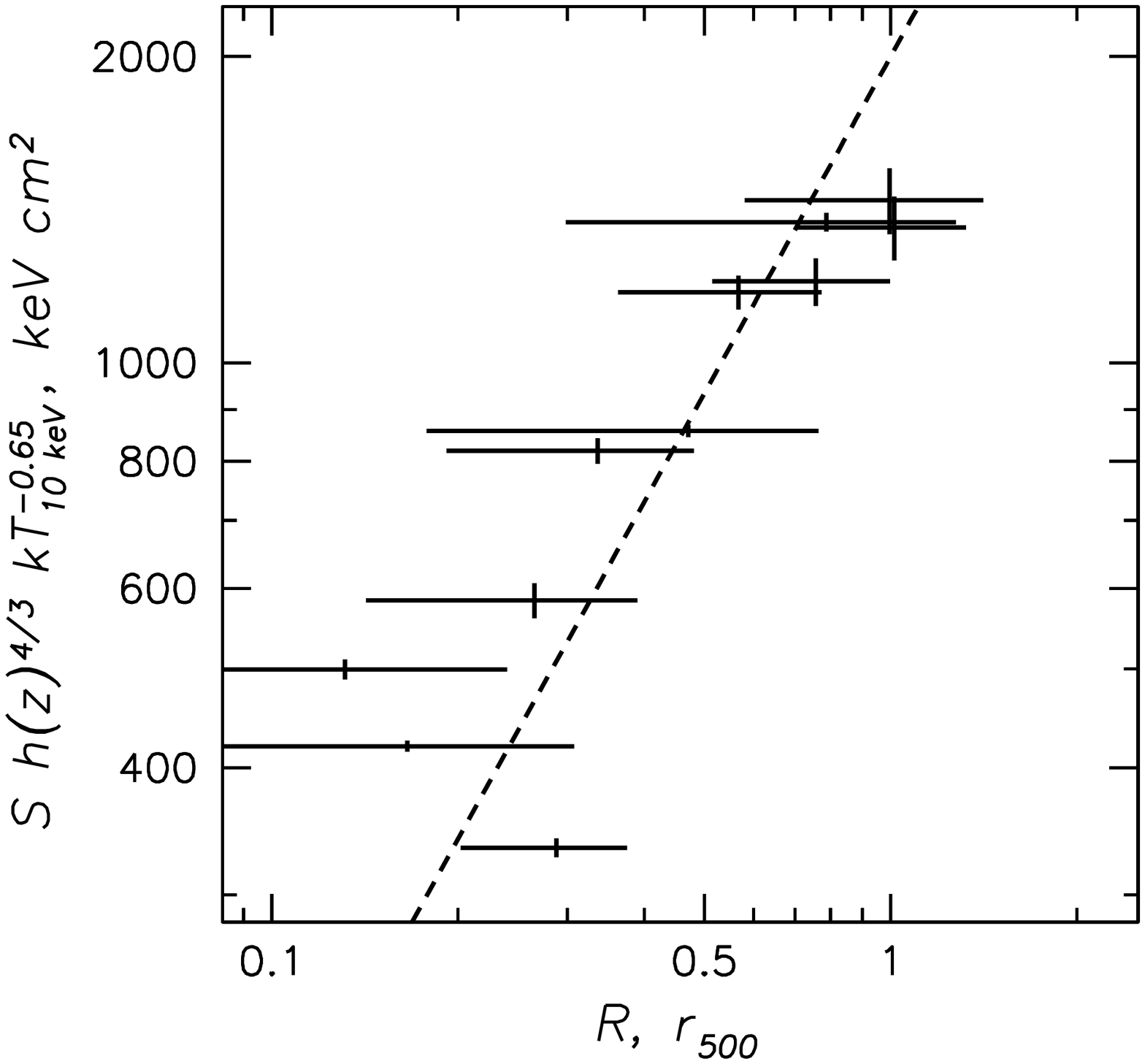}\hfill\includegraphics[width=8.0cm]{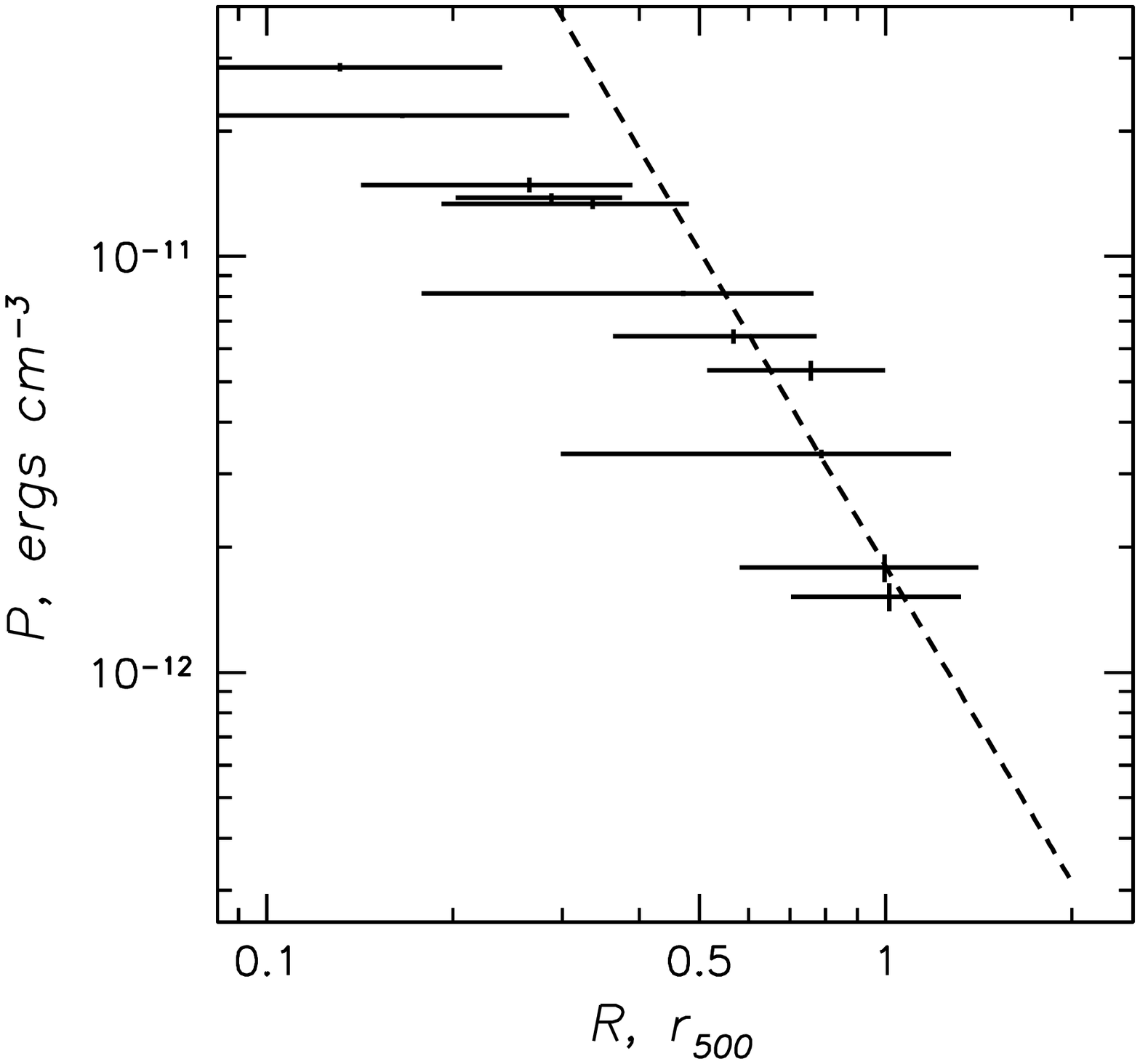}

\caption{Scaled Entropy profile (left panel), and pressure
profile (right panel) of A3667. The universal entropy scaling relation of
Ponman et al. (2003) is shown with a dashed line. A pressure slope of
-2.5, typical for the relaxed clusters is shown with a dashed line.
 \label{f:scale}}
\end{figure*}

Estimation of the projected length has been done assuming a cone geometry,
in which every region is part of a spherical shell that has its inner and
outer radii passing through the nearest and furthest points of the selected
region, respectively. This spherical shell is further intersected by a
cylinder, that is directed towards the observer and in the observer plane
has the cross-section of the selected region. (A more detailed 
description of this estimation is given in Henry et al. (\cite{ph04})). 
We have selected the main pressure peak to be the centre. It also corresponds 
to the position of the brightest galaxy in A3667 and a center of large-scale 
potential revealed by the weak lensing (Joffre et al. \cite{mj00}).

There are a number of features seen in the table. First, the cold front is
the most metal rich zone in A3667. Second, the cold front does not have the
highest pressure. Third, the zones behind the cold front (towards the
north-west) have higher abundance, compared to the bulk of the cluster, yet
lower than of that of the cold front, suggesting that mixing is occurring
there. Fourth, region number 4, the hot tail, is at the same distance from
the pressure peak along the major axis as the cold front. It has a similar
pressure, a factor of two higher temperature, more than a factor of two
higher entropy and lower element abundance compared to the cold front.  But
the gas in region 4 seems to be of the same origin as that in region 5, just
in front of the cold front, because they have the same iron abundance and
the same entropy. The difference in temperatures between regions 4 and 5 
is due to adiabatic expansion of region 5, and indeed the pressures and 
temperatures of the two regions follow the adiabat $T P^{1/\gamma - 1}$ 
= constant.

Let's consider regions 1, 2 and 3, which lie in a line between the pressure
peak and the cold front. Both the entropy and metallicity of region 1 and
the adjacent region 3 are the same and are intermediate between region 2,
the cold front, and the rest of the cluster, regions 4 - 9 
and 11. This suggests
that the material in regions 1 and 3 was previously like the rest of the
cluster but has become polluted by disrupted cold front gas. The original
cold front gas possibly attained its higher metallicity because it was
associated with the brightest cluster galaxy.  If this suggestion is true,
then regions 1 and 3 have been 30\% polluted by cold front gas. The initial
mass of the high metallicity gas that became the cold front is the present
mass of region 2 plus 30\% of the present mass of regions 1 plus 3 or
$5.5\times10^{12}M_\odot$. So the cold front only retains about 30\% of its
original gas (its present mass relative to the initial mass of the high
metallicity gas). About 50\% of the original gas now resides in region 3 and
the remaining 20\% in region 1.

If the gas of the cold front were located at the pressure peak of A3667 its
temperature would be $kT_{\rm cold\; front}\times (P_{\rm center}/P_{\rm cold\;
front})^{\gamma-1 \over \gamma}=5.9\pm0.2$ keV, which is similar to the
temperature of the main component. Heinz et al. (2003) claimed that it is
possible to reproduce the temperature variations in A3667 even having an
isothermal temperature distribution at the beginning. Repeating the same
calculation as for the cold front we conclude that this proposal is
consistent with our data as any temperature variations could be reproduced
via adiabatic compression and expansion, assuming an initial temperature of
5.9 keV. Two caveats of this scenario are: (1) isothermal clusters do 
not show a strong iron abundance gradient (De Grandi \& Molendi \cite{sd01}), 
and (2) the original 5.9 keV temperature, if a virial temperature, implies
a more massive object then a group.

Mazzotta et al. (2002) report Chandra observations of a 300 kpc long
filamentary surface brightness enhancement. The brightest section of this
feature is in our region (10) and it is clearly seen in all four XMM maps:
surface brightness, temperature, entropy and pressure. Mazzotta et
al. interpret this feature, and a similar sized surface brightness
decrement, as a shear (Kelvin-Helmholtz) instability that mixes the colder
gas of the merging object with the hotter ambient material of the main
cluster. Our data provide support for this interpretation. We call region
(10) the shear bar, or bar for short.  Material in the bar seems to
originate from the gas in the core (region 3) because they both have the
same temperature and Fe abundance. Mazzotta et al. (2002) also noted that
the 6 keV gas of the bar is the same as that in the core and is surrounded
by hotter 8 keV ambient gas. New in our data is the agreement of the
abundances. We also mention that the filamentary structure in the entropy
map of the cold front, which we have called the stem of the mushroom, has
similar projected dimensions as that of the bar and so may have similar
origin. More detailed study of the entropy/pressure features associated with
the development of hydrodynamical instabilities in A3667 core will be
presented elsewhere (Finoguenov et al., in preparation).

In Fig.\ref{f:scale} we compare the obtained entropies and pressures with
known scaling relations. In the adopted cosmology, the estimate of the $r_{500}$
is 1.17 Mpc, where the scaling of $r_{500}$ with temperature is taken from
Finoguenov et al. (2001) and the temperature of A3667 is taken as 7.1
keV. As both entropy and $r_{500}$ are scaled by a similar power of
temperature, the deviation of data points in Fig.\ref{f:scale} from the
scaled entropy profile of Ponman et al. (2003) is not very sensitive to the
adopted temperature. As discussed in Ponman et al. (2003), normalization of
the scaled entropy is 30\% different between cold and hot clusters. The
normalization for the hot clusters fits the A3667 data better, thus confirming 
the trend seen in Ponman et al. (2003).

Presence of the disturbed core in A3667 reveals itself in Fig.\ref{f:scale}
as a scatter in the entropy plot, while the pressure is monotonically
decreasing. In the pressure plot region 6 exhibits a slightly higher
pressure and is associated with the dark matter potential of the group. The
regions located at $r_{500}$ exhibit low entropy, this is not very typical
of other clusters and is probably related to the merger scenario of A3667,
where a merging small group brought low-entropy gas with it.

\subsection{Pressure and entropy maps of A3667}

The pressure map appears to be rather symmetric around the
center. However, a strong ellipticity is observed with the major axis
directed towards the cold front. The minor to major axis ratio is
0.68. Beyond 6.8$^\prime$ (430 kpc) north-west from the center of
A3667 along the major axis, the pressure in the tail of A3667 becomes
larger than the pressure the same distance to the south-east, which is
beyond the cold front. This happens roughly between the two main
galaxies of A3667 and could well reflect the potential well of the two
merging components. Alternatively, this asymmetry could be due to pressure
enhancement associated with the turbulence in the tail of the cold
front.  Since the shift of the pressure centroid on even larger scale
of 20$^\prime$ (1.3 Mpc) is still observed, where turbulence is
neither seen nor expected, the first interpretation is favored.

\subsection{New features in the XMM-Newton observations and some
interpretations}

(1) We observe structure in the cold front, best seen in the entropy
maps of Figures 2 and 4. The structure resembles a mushroom head and
stem and bears a close resemblance to the entropy structure at 3 and
4 Gyrs after core passage in the simulation of Heinz et al. (\cite{sh03}).

(2) We find an offset of the mushroom from the pressure peak. Also the
pressure peak coincides with the brightest cluster galaxy. The
pressure map of A3667 exhibits a symmetry around the center, while the
entropy map does not. In combination with (1) this can be interpreted 
as a late stage of the disruption of the core of A3667 by the 
Rayleigh-Taylor instability. Such instabilities are also seen in 
the simulations of Ricker \& Sarazin (2001).

(3) We observed tails of the low-entropy gas coming off the concave
surface of the cold front (to the north-west).

(4) There is an elongation in the pressure and entropy maps to the
north-west from the A3667 center, which we associate with the second
galaxy concentration located there.

(5) In addition to the cold front there is another low temperature feature,
which is located west of the cold front and extends about 600 kpc towards
the north-west, starting at about RA = 20h 12m 25s, DEC = -56$^\circ$
55$^\prime$ 00$^{\prime \prime}$, best seen in the temperature map of Fig. 2
as a green elongated area.  This is a first indication of the deviation in the
cold front from the symmetry in the observer's plane and could be due to
non-negligible angular momentum possessed by the cold-front.

(6) The positional angle of the major axis of the main pressure core is 
different by $15^\circ$ from the positional angle of the major axis of the 
rest of the cluster. The direction from the pressure peak to the center of 
the cold front is the same as the positional angle of the main cluster.

(7) A comparison of the location of the region with minimum entropy in
the XMM data (the cold front) with the sharp edge observed by Chandra
shows that the XMM region is shifted towards the north-east edge of
the cold front (see Fig. 7).  This reinforces the conclusion drawn
from point (5) that the cold front has a velocity component in the
north-east direction.

\begin{figure}
\includegraphics[width=8.5cm]{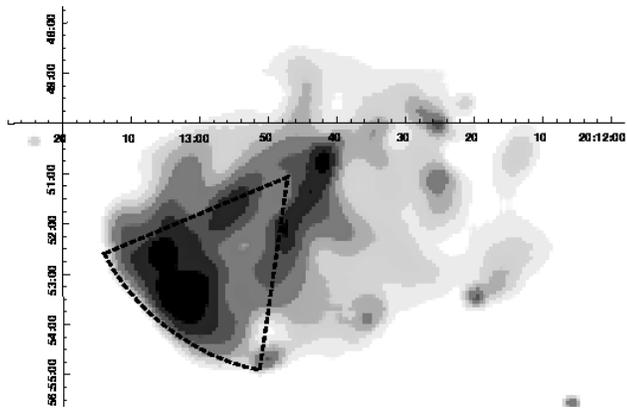}

\caption{
Entropy of the ICM in A3667, as measured by XMM-Newton.
Dashed lines indicate the curvature of the Chandra cold front and
cut out the exact extent of the Chandra cold front (Vikhlinin, A. 2003, private
communication). \label{f:e-ch}}
\end{figure}

\section{Summary}

We have analyzed 6 partially overlapping observations of the nearby galaxy
cluster A3667, done with the XMM-Newton X-ray observatory. We could produce
surface brightness images in different energy bands out to a radius of 21
arcminutes. The large grasp of the observatory enabled us to produce
temperature, pressure and entropy maps done to unprecedented precision. We
confirm the findings of the Chandra observations, especially the cold
front. In addition, we find structure in the cold front, best seen in the
entropy map, which resembles a mushroom shape, very similar to recent
simulations of the dynamics of cold fronts. The pressure map shows symmetry
around its center, which is coincident with the maximum of the surface
brightness map.  We find small scale fluctuations in the inner part of the
pressure map, probably caused by turbulent shock waves. Comparing our
entropy map with the high spatial resolution surface brightness map of the
cold front, obtained with Chandra, we find a significant offset of the
minimum entropy region from the cold front, which we interpret as signs of
angular momentum of the cold front.

%******************************************************************************

\begin{acknowledgements}
The XMM-Newton project is supported by the Bundesministerium f\"ur Bildung
und Forschung/Deutsches Zentrum f\"ur Luft- und Raumfahrt (BMFT/DLR), the
Max-Planck Society and the Heidenhain-Stiftung, and also by PPARC, CEA,
CNES, and ASI. AF acknowledges receiving the Max-Plank-Gesellschaft
Fellowship and support from the Verbundforschung grant 50 OR 0207 of the
DLR.  J. P. Henry thanks Prof. G. Hasinger for the hospitality at the MPE.
We thank the anonymous referee as well as Silvano Molendi and 
Sebastian Heinz for useful comments, which improved the paper.

\end{acknowledgements}

\end{document}